\documentclass[aps,apl,twocolumn,amssymb]{revtex4-1}

\usepackage{graphicx}
\usepackage{dcolumn}
\usepackage{bm}
\usepackage[english]{babel}
\usepackage{color}
\usepackage{amsmath}
\usepackage{graphpap}

\begin{document}

\newcommand{\bise}{Bi$_2$Se$_3$}
\newcommand{\bite}{Bi$_2$Te$_3$}
\newcommand{\sbte}{Sb$_2$Te$_3$}
\newcommand{\bsps}{Bi$_x$Sb$_y$Pb$_z$Se$_3$}
\newcommand{\sro}{ SrTiO$_3$}
\newcommand{\bts}{ Bi$_2$Te$_2$Se}
\newcommand{\bsts}{ Bi$_{2-x}$Sb$_x$Te$_3$Se$_y$}
\newcommand{\rb}[1]{\textcolor{red}{#1}}
\newcommand{\tstern}{\ensuremath{T_2^*} }
\newcommand{\degree}{\ensuremath{^\circ}}
\newcommand*{\fg}[1]{Fig.\thinspace\ref{#1}}
\newcommand*{\fgs}[1]{Figs.\thinspace\ref{#1}}
\newcommand*{\eq}[1]{Eq.\thinspace\ref{#1}}
\newcommand*{\tb}[1]{Tab.\thinspace\ref{#1}}
\newcommand{\todo}[1]{ \textbf{\textcolor{red}{{#1}}}}

%

\title{Phase-coherent transport in catalyst-free vapor phase deposited Bi$_2$Se$_3$ crystals}

%

\author{R. Ockelmann$^{1,2}$, A. M\"{u}ller$^{1}$, J. H. Hwang$^{1}$, S. Jafarpisheh$^{1,2}$, M. Dr\"ogeler$^{1}$, B. Beschoten$^1$ and C. Stampfer$^{1,2}$}

\affiliation{
$^1$JARA-FIT and 2nd Institute of Physics, RWTH Aachen University, 52074 Aachen, Germany, EU \\
$^2$Peter Gr\"unberg Institute (PGI-9), Forschungszentrum J\"ulich, 52425 J\"ulich, Germany, EU
}

\date{\today}

%

\begin{abstract}

Free-standing \bise\ single crystal flakes of variable thickness are grown using a catalyst-free vapor-solid synthesis and are subsequently transferred onto a clean Si$^{++}$/SiO$_2$ substrate where the flakes are contacted in Hall bar geometry. Low temperature magneto-resistance measurements are presented which show a linear magneto-resistance for high magnetic fields and weak anti-localization (WAL) at low fields. Despite an overall strong charge carrier tunability for thinner devices, we find that electron transport is dominated by bulk contributions for all devices. Phase coherence lengths $\l_\phi$ as extracted from WAL measurements increase linearly with increasing electron density exceeding $1~\mu m$ at 1.7~K. While $\l_\phi$ is in qualitative agreement with electron electron interaction-induced dephasing, we find that spin flip scattering processes limit $\l_\phi$ at low temperatures.
\end{abstract}
\maketitle
%
%
\section{INTRODUCTION}
Topological insulators (TIs) are a new class of materials~\cite{bernevig2006quantum,fu2007topological,fu2007topological2,moore2007topological,qi2008topological}, consisting of an insulating bulk and \emph{topologically protected} conducting surface states. These surface states are spin polarised and robust against scattering from non-magnetic impurities making them interesting candidates for future spintronics and quantum computing devices~\cite{loss1998quantum,hasan2010colloquium,pan2011electronic} as well as potential hosts for Majorana fermions~\cite{majorana1937theoria,kitaev2001unpaired,fu2008superconducting,beenakker2013search}. Binary Bi-chalcogenides (\bise , Bi$_2$Te$_3$) belong to the class of three-dimensional (3D) strong topological insulators with a single Dirac cone at the surface~\cite{zhang2009topological} which is experimentally observable by angle-resolved photo emission spectroscopy~\cite{xia2009observation}. In particular, \bise\ with a single Dirac cone centered in a bulk band gap of $E_g\approx 350$~meV is a promising material for probing surface states by electronic transport. However, the measurement of pure surface states is challenging. So far \bise\ crystals are unintentionally $n$-type doped most likely by Se vacancies~\cite{kasparova2005n, scanlon2012controlling, wang2011topological} leading to bulk conductivity dominating electronic transport. To increase the surface-to-bulk ratio, ultra-thin flakes with thickness of the order of 10~nm have been investigated~\cite{yan2013synthesis}.

Thin \bise\ crystals can be produced by mechanical exfoliation of bulk material as it is common practice for graphene fabrication~\cite{novoselov2005two,novoselov2005two-}.
However, it is much more promising to grow thin \bise\ films in-situ which has been successfully achieved with molecular beam epitaxy (MBE)~\cite{zhang2010crossover,zhang2011growth, chen2010gate} or vapor-solid synthesis (VSS) in a tube furnace~\cite{kong2010topological,peng2010aharonov,yan2013synthesis}.

In this work, we show a catalyst free growth of large freestanding \bise\ flakes with a VSS method. Our freestanding growth approach ensures the synthesis of strain free, few-layer single crystal flakes with lateral dimensions up to 25~$\mu$m and thicknesses in the range of 6 to 30~nm, ranking our flakes among the largest \bise\ single crystals flakes. The high structural and surface quality of the  \bise\ crystals is verified by Raman and by scanning force microscopy. The free standing single crystals are  ideal for transport studies. We utilize a wet chemistry-free process which allows transferring these single crystals onto any desired substrate without introducing additional contamination. We studied low temperature magneto-transport on a series of \bise\ crystals of different thicknesses which were transferred on SiO$_2$/Si$^{++}$ substrates. We observe linear magneto-resistance (LMR) at high $B$-fields as well as weak anti-localization (WAL) which both indicate the dominance of bulk transport contributions. Electron phase coherence lengths are in the micrometer range, slightly larger compared to earlier studies on \bise\ crystals grown directly on SiO$_2$~\cite{gao2012gate} or by other growth methods~\cite{alegria2012structural,steinberg2011electrically}. We show that electron spin-flip processes limit the phase coherence length at low temperatures.

\section{CRYSTAL GROWTH AND CHARACTERISATION}

With the goal of gaining high quality thin \bise\ crystals we applied a catalyst-free vapor-solid synthesis method. Most commonly, MBE~\cite{krumrain2011mbe,tarakina2012comparative,borisova2012mode} is used to grow thin films since it offers the growth of extended films of rich chemical compositions with excellent thickness control. Yet it suffers (i) from poly-crystallinity of the film inducing strain at the grain boundaries and (ii) from a limited number of usable substrates. In contrast to MBE, VSS allows the growth of single crystalline platelets on a variety of different substrates~\cite{kong2010few, li2012controlled, gehring2012growth}. However, strain can still be induced by the growth substrate. Moreover, growth catalyst may induce unwanted dopants into the crystal. In this work we therefore use a catalyst-free VSS method where flakes and ribbons grow free standing on the substrate. This offers an interesting pathway for the fabrication of high quality devices. Free standing flakes are neither strained nor contaminated by the substrate material and can be easily transferred onto on a wide range of different substrates.

\begin{figure}[t]%
\includegraphics*[width=\linewidth]{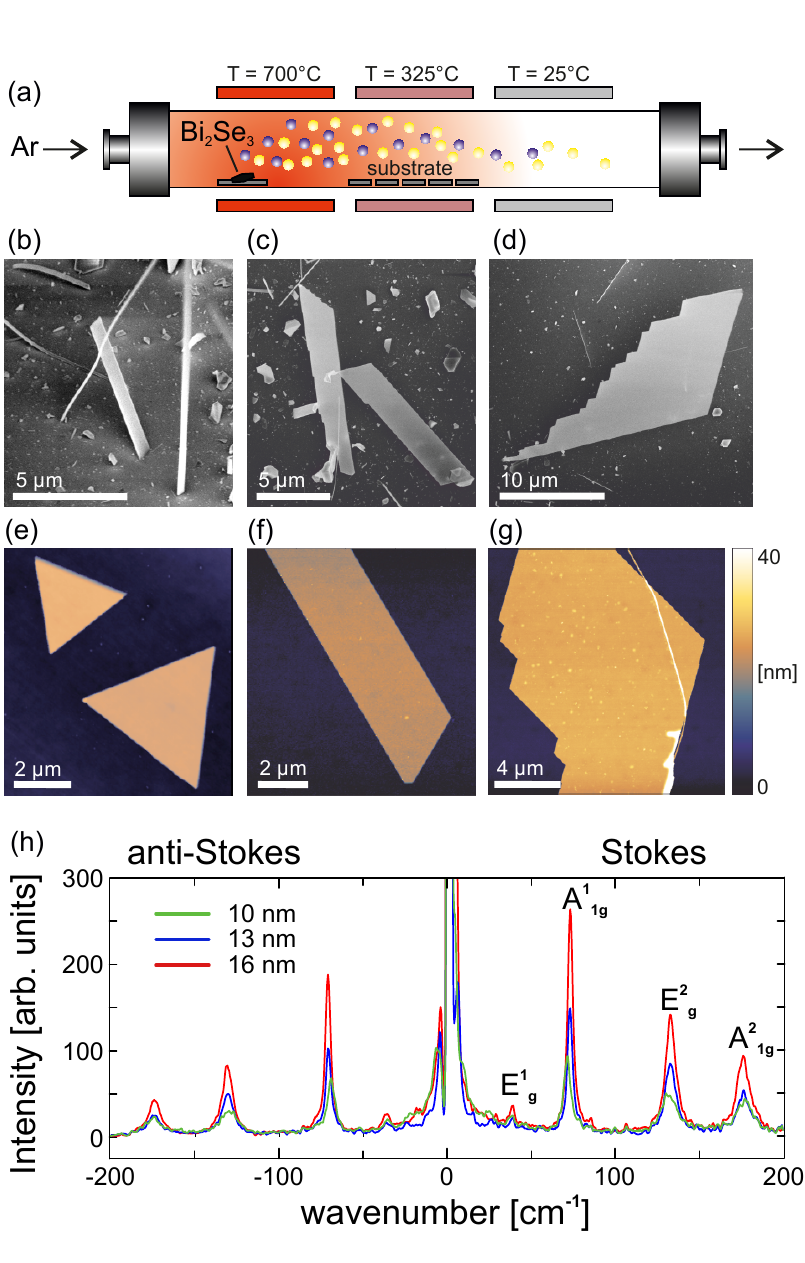}
\caption{\label{fig:figure1}
(color online)
(a) Schematic illustration of the three-zone oven used for \bise\ sample growth. (b), (c) and (d) Scanning electron microscope images of typical 'free standing' \bise\ ribbons and flakes. (e), (f) and (g) AFM images of grown flakes transferred onto SiO$_2$/Si substrate.
(h) Characteristic Raman spectra of grown \bise\ flakes with different thicknesses.}

\end{figure}

A standard three zone tube-furnace (Fig.~\ref{fig:figure1}(a)) with electric heating coils is used for the VSS growth. As a source material we place \bise\ crystals \footnote{\bise\ from Alfa Aesar, Vacuum Deposition Grade, 99.999\%} in the first zone. The Si/SiO$_2$ growth substrates are placed downstream in the second zone. Their exact position was optimized through several growth cycles. Prior to growth, the quartz tube was evacuated to 2~mbar with subsequent argon flushing for 5~min with 500~sccm flow rate which is regulated by a digital mass flow controller. After cleaning, the growth zone (second zone) is heated up to 325\degree C with a constant argon flow of 100~sccm to carry away vaporized particles. The second zone is kept at 325\degree C and 25~mbar for 2~h as it is crucial for the temperature and pressure to be stabilized during growth. Finally, the actual growth process is executed by heating the first zone to 700\degree C. The source material gradually vaporizes and gets carried downstream by an 60~sccm argon flow. Temperature and pressure were optimized to grow large, thin, free standing \bise\ flakes and ribbons as shown by scanning electron microscope (SEM) images in Figs.~\ref{fig:figure1}(b)-(d) and by atomic force microscope (AFM) images in Figs.~\ref{fig:figure1}(e)-(g). Straight edges with only 60\degree\ and 120\degree\ corners indicate single crystalline growth. According to AFM measurements the flake thicknesses range between 6~nm and 40~nm and lateral dimensions can reach up to 25~$\mu$m. AFM images also reveal the flake's surfaces to be stepless confirming a very homogeneous layer by layer growth.

Raman spectroscopy has emerged as an excellent tool to probe crystal stoichiometry of Bi$_2$Se$_3$~\cite{shahil2012micro,tu2014van,guo2010controllable}. The Raman spectra of our Bi$_2$Se$_3$ flakes is obtained using confocal Raman spectroscopy with a laser spot diameter of around 500 nm at a wavelength of 532~nm. The laser spot is precisely positioned on the flakes using a piezo stage. In Fig. 1(h) all four characteristic Raman peaks of Bi$_2$Se$_3$ are clearly seen at 37 cm$^{-1}$, 71 cm$^{-1}$, 131 cm$^{-1}$ and 175 cm$^{-1}$, which correspond to the $E_g^1$, $A_{1g}^1$, $E_g^2$ and $A_{1g}^2$ vibrational modes, respectively. The peak positions are very close to previously measured Raman peaks of stoichiometric Bi$_2$Se$_3$ crystals~\cite{richter1977raman,shahil2012micro,zhang2011raman} indicating the high crystal quality of our flakes.

For transport studies, the free standing \bise\ flakes are dry-transferred by gently dabbing a clean room cloth onto the grown chips and subsequently onto a clean SiO$_2$/Si$^{++}$ substrate.
This method does not involve solvents or other liquids which could effect the surface quality.
These substrates are pre-patterned with gold markers to relocate individual flakes and enable consecutive electron beam lithography (EBL).

\begin{figure}[t]%
\includegraphics*[width=\linewidth]{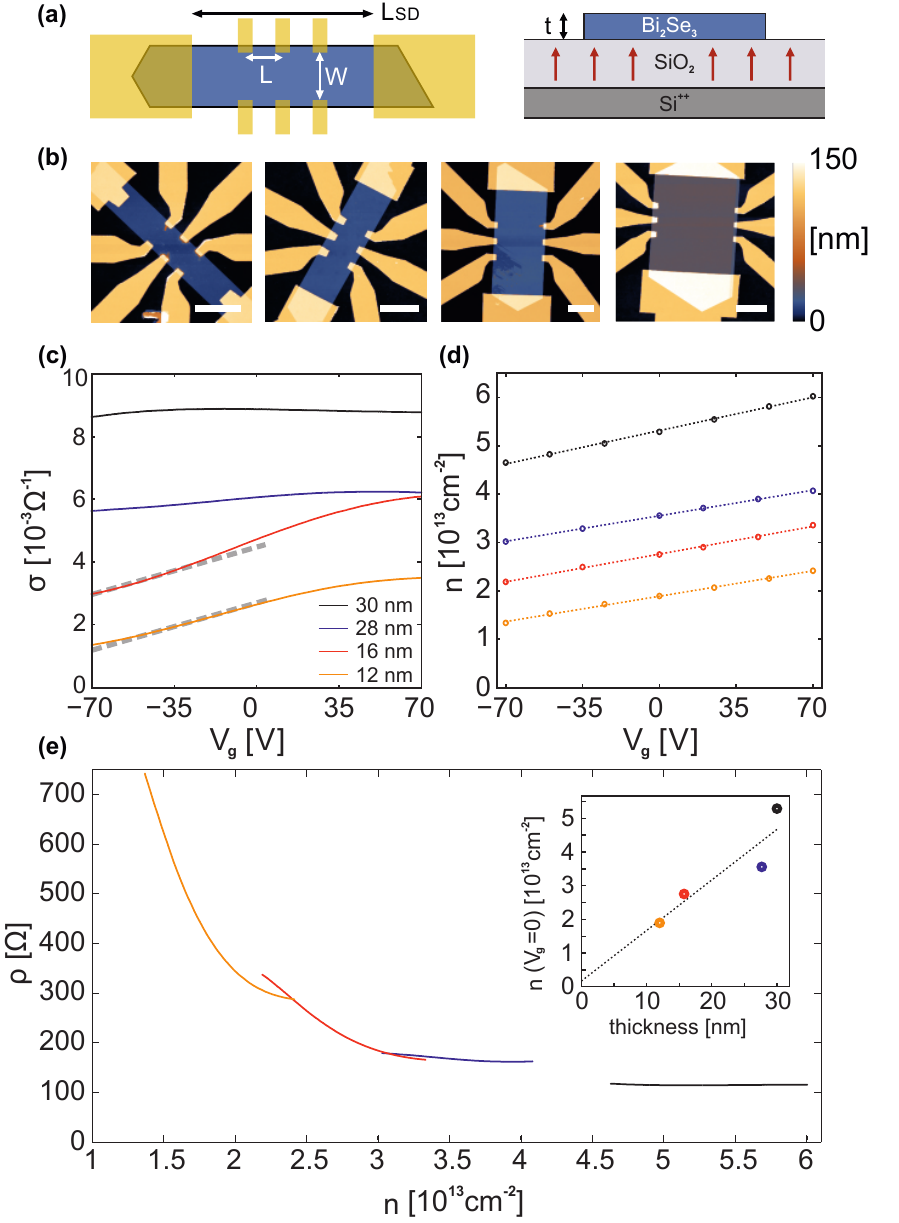}
\caption{\label{fig:figure2}
{  (color online)
(a) Schematic illustrations of the sample geometry of contacted \bise\ flakes. In the left panel we highlight the width ($W$), length ($L$) and source-drain distance ($L_{SD}$) of a contacted sample.
The source-drain distance in our samples is in the range of $14-19~\mu$m. Right panel shows a cross-section of our samples highlighting the flake with thickness ($t$) resting on SiO$_2$ with back gate. (b) AFM images of the four investigated devices contacted with electrical contacts for Hall effect and magneto-resistance measurements. The scale bars are all 5~$\mu$m. For more details on the geometry please see Table \ref{tab:dimensions}. (c) Conductivity as function of back gate voltage for all four samples.
(d) Two-dimensional carrier density $n$ of the different samples, extracted from the Hall resistance vs back gate voltages. (e) Resistivity as function of carrier density for all devices. The inset shows the carrier density at zero back gate voltage as function of \bise\ flake thickness.
}
}
\end{figure}

The flakes to be contacted are first chosen by optical microscopy and further characterized by AFM, which is also used to determine exact dimensions. The contacts are defined using standard EBL techniques and 5/50~nm Cr/Au ohmic contacts. Directly before metal evaporation the contact areas are etched for 15~s by oxygen plasma to remove any oxide layer from the \bise\ surface. This step is crucial for low contact resistances. The contact geometry of flakes with a high length/width ratio, resembles a fairly good approximation of a Hall bar geometry (\textit{cf.} Figs.~\ref{fig:figure2}(a),(b)). With this method no additional patterning step is needed, allowing us to keep unetched flake edges, as grown in the VSS-process. The dimensions of the four investigated devices are summarized in Table~\ref{tab:dimensions}.

\begin{table}[b]
\begin{tabular}{@{}ccllc@{}}
 \vspace{0.05 in}
     device \# \quad & $t$~(nm) \quad & $L$~($\mu$m) \quad & $W$~($\mu$m) \quad & $\alpha_g$~(cm$^{-2}$V$^{-1}$) \\
    \hline
		\hline \vspace{-0.1 in} \\
		1 & 12  & 3.4 & 2.4 & 7.5$\times10^{10}$ \\
		2 & 16  & 3.5 & 2.9 & 8.2$\times10^{10}$ \\
    3 & 28  & 4.7 & 10.6 & 7.5$\times10^{10}$ \\
    4 & 30  & 3.5 & 11.8 & 9.8$\times10^{10}$ \\

    \hline
\end{tabular}
\caption{\label{tab:dimensions} Geometrical dimensions of the four devices discussed. The dimensions are defined as in Fig.~\ref{fig:figure2}(a) and were measured using an AFM. The respective gate lever arms $\alpha_g$ are also included. }
\end{table}

\begin{figure}[t]%
\includegraphics*[width=\linewidth]{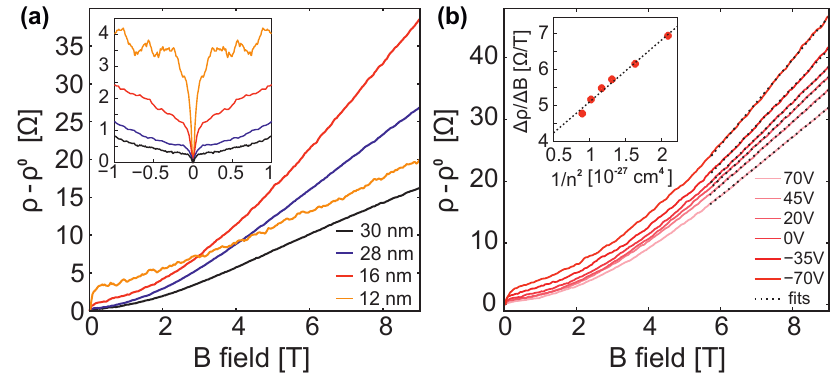}
\caption{\label{fig:figure3}
{  (color online)
(a)
Magneto-resistance $\Delta \rho = \rho-\rho^{0}$ [where $\rho^{0}=\rho(B=0~T)$] at zero back gate voltage for the four devices in Fig.~2. The inset shows WAL dips at small $B$ fields. (b) $\Delta \rho$  as function of $B$-field for the 16~nm thick sample for various back gate voltages (see labels).
The inset shows the slopes of the linear magneto-resistance at high magnetic fields (see dashed lines in main panel) vs $1/n^2$.
}
}
\end{figure}
\begin{figure*}[t]%
\includegraphics*[width=\linewidth]{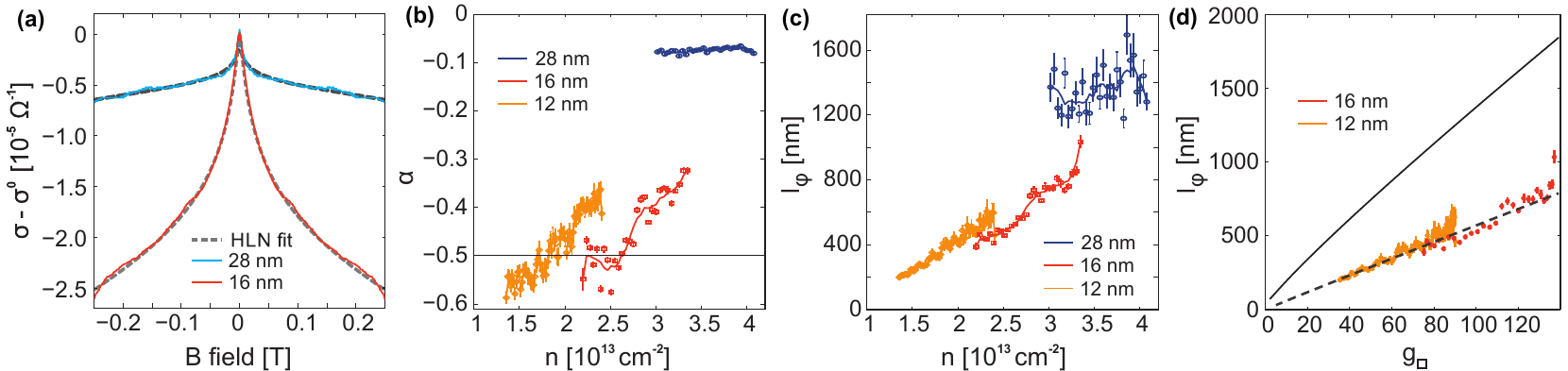}
\caption{ \label{fig:figure4}
{  (color online)
Weak antilocalization (WAL) peak in conductivity as function of $B$ field (solid lines) and fits according the HLN model (dashed lines) for two different devices. (b), (c) Extracted fitting values for $\alpha$ (panel b) and $l_{\phi}$ (panel c) are shown as a function of the charge carrier density for three different devices.
(d) Phase coherence length as function of the dimensionless conductivity $g_{_{\square}}$. Here, the solid line displays the theoretical result by AAK~\cite{AAK}. The dashed line highlights
the result modified due to a finite electron spin-flip scattering time $\tau_{sf}$ (see text for details).
}
}
\end{figure*}

\section{RESULTS AND DISCUSSION}

Transport measurements were performed in a $^4$He-cryostat at a base temperature of $T$~=~1.7~K using low-frequency lock-in techniques. A superconducting solenoid, immersed in liquid Helium was used to apply magnetic fields perpendicular to the sample plane.
The back gate characteristics of four different devices (see Fig.~2(b)) with different Bi$_2$Se$_3$ crystal thickness are shown in Fig.~2(c) which depicts the four-terminal conductivity $\sigma$ as function of applied back gate voltage $V_g$. For the 28 and 30~nm thick Bi$_2$Se$_3$ samples almost no gate tunability is observed, which is in contrast to the two thinner (12~nm and  16~nm thick) samples where $\sigma$ can be tuned by a factor of around 2. In none of our samples we observe an ambipolar transport behavior, which is a first
indication that very high n-doping of \bise~is present in all our devices.

We performed Hall effect measurements to determine the charge carrier densities. The extracted two-dimensional (2D) electron density, $n$, which varies linearly with $V_g$ is shown in Fig.~2(d) for all devices. The slope for the three thinner samples of $7.5-8.2\times 10^{10}$~cm$^{-2}$/V (see table~1) is in reasonably good agreement with the geometrical gate lever arm $\alpha_g=\epsilon_0 \epsilon_r /(|e|d) \approx$~7.2~$\times 10^{10}$~cm$^{-2}$/V, where $d = 285$~nm is the thickness of the SiO$_2$ gate oxide with a dielectric constant of $\epsilon_r \approx$~4.
From the vertical offsets of $n$ in Fig.~2(d) we can estimate the average
bulk charge carrier density.
The inset in Fig.~2(e) shows the 2D carrier density $n$ at zero gate voltage as function of the
Bi$_2$Se$_3$ crystal thickness. From the slope of this linear dependence (dotted line) we extract a 3D bulk
carrier concentration of 1.5$\times 10^{19}$~cm$^{-3}$ in our VSS Bi$_2$Se$_3$ crystals.
Finally, we plot the resistivity $\rho=1/\sigma$ of all four devices as function of $n$
(main panel of Fig.~2(e)). The observed overall
trend highlights a consistent carrier density dependency on the measured resistivity of all measured devices.
The increasing gate tunability at lower carrier densities
might be either connected (i) to the linear density of states
of the 2D surface states or (ii) to a reduction of the 3D bulk density of states (or diffusion constant) at lower Fermi energy values.
By assuming that surface states dominate the gate voltage dependence at low carrier densities we can estimate the carrier mobility from the nearly linear increase of $\sigma$
as function of $V_g$ (see dashed lines in Fig.~2(c)).
From our data we extract respective surface carrier mobilities of $\mu \approx 1600$~cm$^2$(Vs)$^{-1}$.
As the condition $\mu B > 1$ can be reached with experimentally accessible magnetic fields, Shubnikov-de Haas oscillations
should become visible for $B$-fields larger than 5~T.
We note, however, that we do not observe any Shubnikov-de Haas oscillations
(see magneto-transport measurements below).
We therefore conclude that is very likely that transport in our samples is dominated by bulk transport, where the gate tunability originates from a Fermi level dependent 3D bulk density of states or diffusion constant.

For gaining more insights on the separation of bulk and surface transport we performed
four-terminal magneto-resistance measurements (see Fig.~3(a)).
The two most prominent features in our magneto-transport data are (i) a LMR at high magnetic fields~\cite{tang2011two,he2012high,chiu2013weak, yan2013large} and (ii) a reasonably strong WAL dip at low magnetic fields (below 1~T)~\cite{lu2011weak,lu2011competition,kim2011thickness,kim2013coherent,chiu2013weak}, as shown by the inset of Fig.~3(a).

Above $B=5-6$~T all four devices exhibit LMR.
Interestingly, the strength of the LMR does not solely depend
on the sample thickness nor the total charge carrier density.
We observe that the thinnest and thickest Bi$_2$Se$_3$ crystal exhibit similar LMR slopes (see black and orange curves in Fig.~3(a)), whereas the other two devices (see red and blue curves in Fig.~3(a)) show a significantly larger slope of the LMR.
However, within a single device we find a systematic carrier
density dependence of the slope of the LMR at large magnetic fields (see Fig.~3(b)).
A more detailed analysis of the carrier-density dependent LMR slope shows that,
interestingly, $\Delta \rho/\Delta B$ changes linearly as function of $n^{-2}$ (see inset in Fig.~3(b)).
This dependence is in agreement with the generic quantum description of galvanomagnetic phenomena by
Abrikosov~\cite{abrikosov1969galvanomagnetic,abrikosov2003,hu2008classical}, leading to $\rho \propto B/n^2$.
However, although for the investigated $B$ field range the required condition $\mu B > 1$ might be fulfilled, we are certainly
not in the extreme quantum limit where only the lowest Landau level is filled. Moreover, it should be noted
that the fit shown in the inset of Fig.~3(b) does not cross the origin. All these bring us to the conclusion that the LMR in our devices is rather dominated by the classical linear  magneto-resistance~\cite{hu2008classical} due to bulk inhomogeneities and defects in the Bi$_2$Se$_3$ crystals which may also explain the high bulk carrier densities.

These findings are in contrast to the WAL, which exhibits a clear crystal thickness dependence (see inset of Fig.~3(a)) and
which is therefore - also in agreement with literature ~\cite{kim2011thickness,kim2013coherent,bansal2012thickness} - most likely a better fingerprint for  surface state transport.
WAL signatures are indeed inherent to the 2D states of TIs ~\cite{chen2010gate,fu2007topological2,lu2011weak}
As it is governed by quantum mechanical interference, a detailed investigation
of resulting corrections to the conductance allows to learn more about
phase coherent transport properties in these materials.
Indeed, WAL in TIs has already been studied in great detail ~\cite{taskin2012manifestation,checkelsky2011bulk,he2011impurity,cha2012weak}  and it has been shown that
the so-called
Hikami-Larkin-Nagaoka (HLN) model~\cite{hikami1980spin} can
be used to fit the WAL corrections at low $B$ fields.
Within the HLN model the conductivity correction is expressed as
\begin{equation}
\begin{split}
\label{eq:HLN}
&\Delta \sigma =\sigma\left(B\right)-\sigma\left(0\right) = \\
&- \frac{\alpha e^2}{2 \pi^2 \hbar} \left[ \ln\left( \frac{\hbar}{4 B e l_\phi} \right) - \Psi\left(\frac{1}{2}+ \frac{\hbar}{4 B e l_\phi} \right)\right],
\end{split}
\end{equation}
where $\Psi$ is the digamma function and
$l_\phi$ is the phase coherence length. The value of the amplitude $\alpha$ is expected to be $-1/2$ for perfect WAL in a two-dimensional system. For an ideal 3D TI with two independent and decoupled 2D surfaces the expected value for the total WAL amplitude is therefore $\alpha = -1$~\cite{kim2013coherent}. For fitting our data, the symmetric and anti-symmetric part of the overall conductivity were separated. This is necessary considering the imperfect Hall bar geometry due to the etch-free sample fabrication process. Fig.~4(a) shows WAL data, fitted with the HLN model given by Eq.~(\ref{eq:HLN}) for the symmetric part of the data with an additional term for quadratic magneto-resistance at low magnetic fields $\beta B^2$. The values for $\alpha$ and $l_\phi$ as extracted from the fits are shown in Figs.~4(b) and 4(c).\\

\begin{figure*}[]%
\includegraphics*[width=\linewidth]{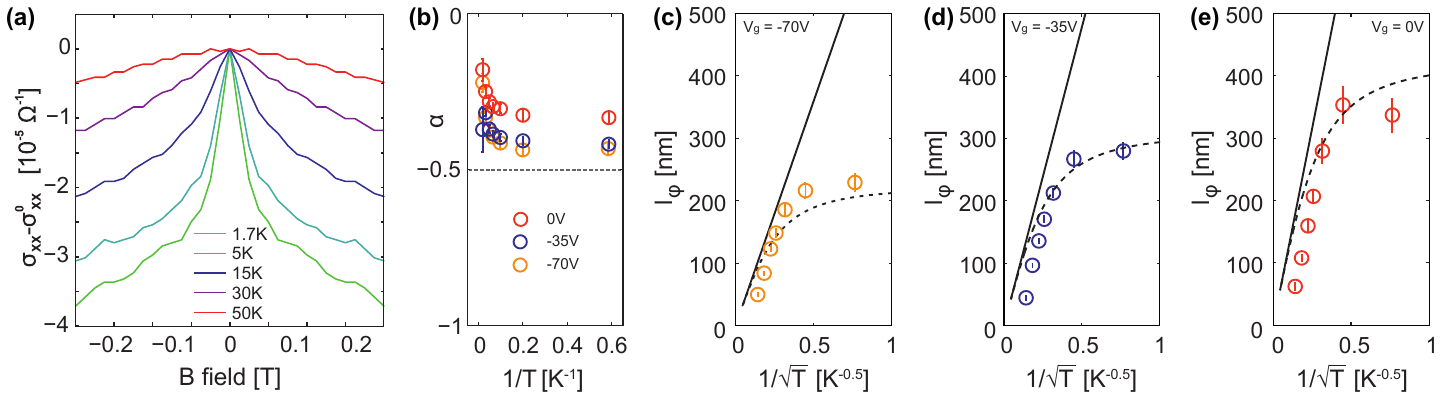}
\caption{ \label{fig:figure5}
{ (color online)
(a) Broadening of the WAL peak with increasing temperature for the 16~nm thick \bise\ sample.
(b) Dependence of the parameter $\alpha$ as a function of inverse temperature for different back grate voltages.
(c),(d) and (e) Dependence of the phase coherence length $l_{\phi}$ as a function of $1/\sqrt{T}$ for three different back gate voltages. The solid and dashed lines resemble the same theoretical models as in Fig. 4(d).}
}
\end{figure*}

For the two thinner samples (orange and red data in Fig.~4(b))
$\alpha$ is gate tunable around a value of $-1/2$.
This indicates that either the surface states are strongly
coupled via the highly conductive bulk or that
transport is purely dominated by the bulk.
For the 28~nm thick sample (blue data in Figs.~4(b), 4(c)) no
gate dependence is observed indicating the dominance of a bulk conduction channel with a
Fermi level in a regime with
constant 3D bulk density of state which suppresses any gate tunability.
The thickest sample (30~nm) does not show a distinct WAL peak and is hence excluded from our WAL analysis. A similar trend is also observed for the phase coherence length $l_\phi$ which increases with larger sample thickness and increasing charge carrier density.

Interestingly, such a gate-tunable
phase-coherence length - also observed by other groups ~\cite{steinberg2011electrically,jauregui2014gate} - is in qualitative agreement with a scattering mechanism based
on electron-electron interactions as predicted by Altshuler-Aronov-Khmelnitsky (AAK) for a two-dimensional system ~\cite{AAK,engels2014limitations}:
\begin{equation}
\label{eq:AAK}
\quad l_{\phi} = \hbar g_{_{\square}} \left(4m^{*} k_B T \;\ln g_{_{\square}}\right)^{-1/2},
\end{equation}
where $k_B$ is the Boltzmann
constant, $m^{*}$ is the effective mass and $g_{_{\square}}=\sigma h/e^2$ is the dimensionless conductivity, which can be directly extracted from the
measured conductivity.
Apart from the small logarithmic correction (only becoming important for very small conductivities)
the phase coherence length is a linear function of $g_{_{\square}}$.
By plotting the experimentally extracted $l_{\phi}$ as function of $g_{_{\square}}$
(see Fig. 4(d)) we indeed can confirm this nearly linear dependence.
Furthermore, by assuming a bulk carrier effective mass of $m^* = 0.15 m_e$ ~\cite{orlita2015magneto},
we obtain the solid line in Fig. 4(d) without any further adjustable
parameter. These values
are a factor of $2-3$ larger than the values of $l_{\phi}$ extracted from
our WAL measurements, meaning that there must be some corrections to
the effective mass or (more likely) additional sources for dephasing.

This becomes even more
apparent when investigating the temperature dependence of the WAL and
the extracted $l_{\phi}$ at different carrier densities, as shown in
Fig.~5.
In Fig.~5(a) we show the WAL peak for the 16~nm thick sample
at different temperatures, highlighting its disappearing
at elevated temperatures. The peak at small magnetic fields slowly decreases in amplitude completely disappears at 50~K.
By fitting again the HNL model to our data we extract the temperature
dependent $\alpha$ values and phase-coherence lengths (Figs.~5(b)-5(e)).
The prefactor $\alpha$ changes towards zero for increasing temperature, i.e. decreasing $1/T$,
as seen in Fig.~5(b). More insights can be gained when investigating the temperature dependence of the
phase coherence length.
In order to highlight the expected temperature dependence given
by Eq.~2 we plot $l_{\phi}$ as function of $1/\sqrt{T}$ in Figs.~5(c) to 5(e) .

Similar to Fig. 4(d), the solid lines
illustrate the estimates for $l_{\phi}$ obtained from the AAK theory
($l_{\phi} \propto 1/\sqrt{T}$) for different back gate voltages, i.e. carrier
densities which are color coded in Figs. 5(c)-5(e).
Indeed, above T = 7 K (below $1/\sqrt{T} \approx 0.4$~K$^{-1/2}$), the
experimentally extracted $l_{\phi}$ values are inversely proportional to the
square root of the temperature. However, at lower temperatures, $l_{\phi}$ shows a
carrier density dependent saturation behavior, which can not be explained by
electron-electron interaction limiting the phase coherence length.
To account for these discrepancies, we
follow Ref. ~\cite{engels2014limitations} and include an additional inelastic electron
spin-flip scattering time, $\tau_{sf}$.
Thus the phase coherence time $\tau_{\phi}=l_{\phi}^2/D$
will be limited by $\tau_{sf}$ at low temperatures.
This leads to an overall scattering rate which is the
sum of spin flip and the AAK decoherence rate, $\tau_{\phi}^{-1} = \tau_{sf}^{-1}+k_B T \ln g_{_{\square}}/\left(\hbar g_{_{\square}}\right)$.
We use this expression to estimate $l_{\phi}=\sqrt{D \tau_{\phi}}$.
By assuming a linear carrier density dependency of
 the spin flip scattering time $\tau_{sf}=\beta n$ with $\beta=1.2 \times 10^{-24}$~cm$^2$s,
we obtain good agreement with all our experimental data (see dashed lines in Fig. 4(d) and
Figs.~5(c) to (e)).
The extracted $\tau_{sf}$ values are on the order of 10~ps. This is a first experimental estimate of the spin flip scattering time in \bise~from WAL data. The short timescale most likely results from the strong spin-orbit interaction in this material class~\cite{Pesin2012}. We emphasize that the $\tau_s$ value is not attributed to surface but rather to bulk transport properties.


%
%
%
\section{CONCLUSION}

In conclusion, we used a catalyst-free vapor solid synthesis growth method for obtaining
well-shaped single-crystalline \bise\ flakes with thicknesses in the range of a few nanometers.
We performed low-temperature transport measurements on such \bise\ flakes with different layer
thicknesses, resulting in different (two-dimensional) doping values.
From magneto-transport measurements we extract information on the linear magneto-resistance
as well as on phase coherent transport properties.
In particular from weak antilocalization measurements we gain detailed insights on the phase-coherence length in bulk transport.
We observe that the phase-coherence length linearly depends on both the conductivity and electron density. Its values are close to the values imposed by electron-electron interaction but limited by spin-flip scattering at the lowest temperatures.

%
\subsection*{Acknowledgments}

We gratefully acknowledge support from the Helmholtz nanoelectronic facility (HNF), the Helmholtz Virtual Institute of Topological insulators (VITI) and the DFG priority program SPP1666.

\newpage

\addtocontents{toc}{\vspace{.5\baselineskip}}
\addcontentsline{toc}{section}{\protect\numberline{}{References}}

\bibliography{paper}
\bibliographystyle{ieeetr}
\end{document}